\documentclass[twocolumn,floatfix]{aastex631}

\usepackage{hyperref}
\begin{document}

\title{\textit{AstroSat} View of the Neutron Star Low-mass X-Ray Binary GX 5-1}

\email{shyamvp151@gmail.com}

\author{Shyam Prakash V P}
\affiliation{Space Astronomy Group, ISITE Campus, U. R. Rao Satellite Center,\\ ISRO, Bengaluru 560037, India}
\affiliation{Department of Physics, University of Calicut, Kerala 673635, India}

\author{Vivek K. Agrawal}
\affiliation{Space Astronomy Group, ISITE Campus, U. R. Rao Satellite Center,\\ ISRO, Bengaluru 560037, India}



\begin{abstract}
We present the spectral and timing study of the bright Neutron Star low mass X-ray binary ((NS-LMXB)) GX 5-1 using \textit{AstroSat/LAXPC} and \textit{SXT} observations conducted in 2018. During the observation, the source traces out the complete horizontal branch (HB) and normal branch (NB) of the Z-track in the hardness-intensity diagram (HID). Understanding the spectral and temporal evolution of the source along the 'Z' track can probe the accretion process in the vicinity of a neutron star. Spectral analysis was performed in the 0.7–20 keV energy range for different segments in the HID using a multi-temperature disc black body with an average temperature, $kT_{in} \sim$0.47, and a thermal Comptonization model. It is found that the optical depth of the corona drops from $\sim$7.07 in HB to $\sim$2.61 in NB. The timing analysis using the LAXPC instrument indicates the presence of quasi-periodic oscillations in HB and NB of the Z-track. The observed QPO frequencies are similar to the characteristic frequencies of horizontal branch oscillations (HBO) and normal branch oscillations (NBO). The HBO frequency increases from $\sim$12-40 Hz towards the hard apex. NBO are observed at $\sim$5 Hz. The timing studies conducted in soft and hard bands indicate the association of HBO and NBO origin with the non-thermal component. Further research could explore the implications of this relationship for understanding the dynamics of accretion onto neutron stars. 


\end{abstract}



\section{Introduction} \label{sec:intro}

X-ray binaries are a class of objects that consist of a compact object, a black hole or a neutron star, and an optical companion star (\cite{1973A&A....24..337S}) that emit X-ray radiation. Depending on the ratio of X-ray to optical luminosity and the mass of the companion star, X-ray binaries can be classified into low-mass X-ray binaries (LMXBs) and high-mass X-ray binaries (HMXBs). In neutron star LMXBs, the compact object is a neutron star, and the mass of the companion star is less than or equal to solar mass. In such a binary system, neutron stars accrete matter from a companion star by Roche-lobe overflow through the inner Lagrangian point (\cite{2006csxs.book..623T}). They have a relatively low surface magnetic field and spin periods compared to other binary systems. LMXBs are further classified into “Z” and “Atoll” sources based on the patterns they trace in the Hardness Intensity Diagram (HID) and Color-Color Diagram (CCD) (\cite{Hasinger_Vanderklis_1989}). An 'Atoll' source traces a banana and island state in its HID. In Z sources, a change in the mass accretion rate is considered to be the reason for tracing the Z pattern in the HID (\cite{Hasinger_Vanderklis_1989}, \cite{Muno_2003}, \cite{Homan_2007}).

\vspace{0.2cm}
Various models were proposed to explain the continuum spectrum of LMXB systems. One of the most important is the Eastern model, which comprises a multi-color blackbody emission from the accretion disc and a Comptonized emission from the close regions of the neutron star (\cite{1989PASJ...41...97M}). The Western model (\cite{White_1988}) suggests a black body emission component from a boundary layer present between the neutron star and the accretion disc. Both the Eastern and Western models take into account the hot plasma that can produce a hard X-ray spectrum. Reflection features, if present in the spectra (e.g., \cite{reflection_ionized}), can provide more information on the changes in accretion flow along the Z-track. An Fe K$_{\alpha}$ line around 6.4 keV due to fluorescence (\cite{Asai_2000}) and a reflection hump observed between 10–30 keV (\cite{2007ApJ...657..448F}) due to Compton back-scattering can be present. Both of these features result from the irradiation of the inner accretion disc and are sensitive to changes in accretion geometry. \cite{Homan_2018} has discussed the absence of the K$_{\alpha}$ line in the NuSTAR spectrum of GX 5-1.

\vspace{0.2cm}
Quasi-periodic oscillations (QPO) are observed as peaks in the power density spectrum (PDS) of the source, which is characterized both by a centroid frequency and its width. The properties of these oscillations shown by the source are highly dependent on the position within the HID. QPOs observed in the HB are termed horizontal branch oscillations (HBOs). It has a characteristic frequency in the range of 15–60 Hz (\cite{Wijnands_1998}). Normal Branch Oscillations (NBOs) are seen in the normal branch of the Z-track at a much shorter frequency (5-7 Hz) (\cite{vanderklis2004review}). Apart from these branch oscillations, millisecond and kilohertz QPOs have also been detected in many LMXBs (see \cite{2000ARA&A..38..717V}, \cite{Wijnands_1998}, \cite{vanderklis_2000}, \cite{Jonker_2000}), with most of the kilohertz QPOs displaying twin peaks. There have been many attempts in the past to explain the origin of these features, but it is still unclear.

\vspace{0.2cm}
GX 5-1 is a persistent neutron star LMXB, discovered in 1968 (\cite{1968ApJ...151....1F}; \cite{1968ApJ...152.1005B}). Further, it is classified as a “Z” source as it traces a Z-pattern in its HID (\cite{1989A&A...225...79H}). It is the second-brightest persistent low-mass X-ray binary located in the direction of the galactic center. It has luminosity close to Eddington luminosity (L/L$_{Edd}$) = 1.6 - 2.3). It is also considered a Cygnus-like source because it shows a short flaring branch (FB) and an extended horizontal branch (\cite{Cyg_like}). The source has exhibited NBO and HBO along with kHz QPOs in the past (\cite{1985Natur.316..225V}, \cite{Wijnands_1998}). A model for Z-track in the source was studied earlier using RXTE data by \cite{refId0}. The variable energy-dependent X-ray polarization in the source has been studied by \cite{fabiani2023discovery}. \cite{Bhulla} has studied the source using \textit{AstroSat} observations conducted in the year 2017. Anti-correlated soft and hard lags were detected in the source using \textit{AstroSat} data by \cite{2022MNRAS.516.2500C}. A review by \cite{polarised_XRBs} explains that the polarization signatures from the source are highly correlated with the position in the HID, with maximum polarization in the HB.

\vspace{0.2cm}
Even decades after the first studies of the evolution of the Z-track in LMXB sources, the cause of this evolution is not clearly understood. In this work, we attempt a broadband spectral and timing study of the Cyg-like Z-source GX 5-1 using \textit{AstroSat} data. The observations cover the entire HB and NB of the Z-track. The broadband spectral capability and good timing resolution of the LAXPC instrument on \textit{AstroSat} help us investigate the spectral and temporal evolution of the source throughout different states along the Z-track.

\vspace{0.2cm}
The paper is organized as follows: The details of the observations and the data reduction are provided in Section \ref{sec:data_reduction}. Details of the spectral and temporal analyses are presented in Section \ref{sec:analysis}. The observations and findings from the analysis are given in Section \ref{sec: results and discussion}. Finally, we discuss the results from the present study in Section \ref{sec:summary}.
 
\section{Observation and Data reduction} \label{sec:data_reduction}

\textit{\textit{AstroSat}} has observed the LMXB source, GX 5-1, several times from February 2018 to May 2018 using the LAXPC (\cite{Antia_et_al}) and SXT (\cite{SXT}) instruments with observation IDs \texttt{G08\_068T01\_9000001922} and \texttt{G08\_068T01\_9000002090} with an effective exposure time of $\sim$42 ks. Observations were conducted for LAXPC in event mode and for SXT in photon counting mode. Simultaneous observational data from these instruments is used for spectral analysis. Out of the three LAXPC instruments, LAXPC10 and LAXPC30 exhibited anomalous behaviors, including gain change and gas leakage. For this investigation, we used archived data from the LAXPC20 instrument. The source traces most of the "Z" track during the observation, except the flaring branch (FB). We reduced the LAXPC data using \href{https://www.tifr.res.in/~AstroSat_laxpc/LaxpcSoft.html} {\texttt{laxpcsoftv3.4.3\_07May2022}}\footnote{\href{https://www.tifr.res.in/astrosat_laxpc/software.html}{https://www.tifr.res.in/astrosat\_laxpc/software.html}} developed by the Tata Institute of Fundamental Research (TIFR). Level 1 data is converted to Level 2 using the software for generating scientific products. The software also includes calibration files and responses. The GTI file is generated, and the light curves and spectra are extracted from the Level-2 files.
     
\begin{figure}
    \centering
    \includegraphics[scale=0.45]{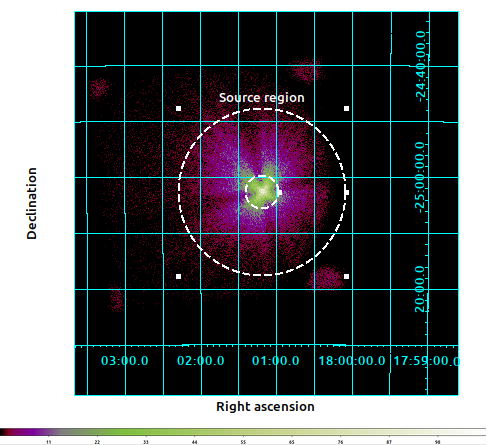}
    \caption{SXT image of the source in the 0.7 – 7 keV energy band. The annular region used to extract the source counts to generate light curves and spectra is also shown.}
    \label{SXT_image}
\end{figure}

\vspace{0.2cm}
SXT data reduction was carried out using the SXT software. Photon counting mode data were merged using the SXT Event Merger Tool\footnote{\href{https://www.tifr.res.in/astrosat_sxt/dataanalysis.html}{https://www.tifr.res.in/astrosat\_sxt/dataanalysis.html}} and used the merged event file to extract the image. The source counts were extracted from a circular region of radius 15 arcmins, and a region of radius 3 arcmins was excluded from the center to avoid pile-up issues. The SXT region selection for the extraction of source counts is shown in Figure \ref{SXT_image}. The latest available RMF (sxt\_pc\_mat\_g0to12.rmf) and background (SkyBkg\_comb\_EL3p5\_Cl\_Rd16p0\_v01.pha) files were used during analysis. The ARF file was created using the SXTARFModule tool\footnote{\href{https://www.tifr.res.in/~AstroSat_sxt/dataanalysis.html}{https://www.tifr.res.in/\~AstroSat\_sxt/dataanalysis.html}} after correcting for the vignetting effect.

\section{Analysis} \label{sec:analysis}
\subsection{Hardness-Intensity Diagram} \label{HID_sect}
\begin{figure}
    \centering
    \includegraphics[scale=0.5]{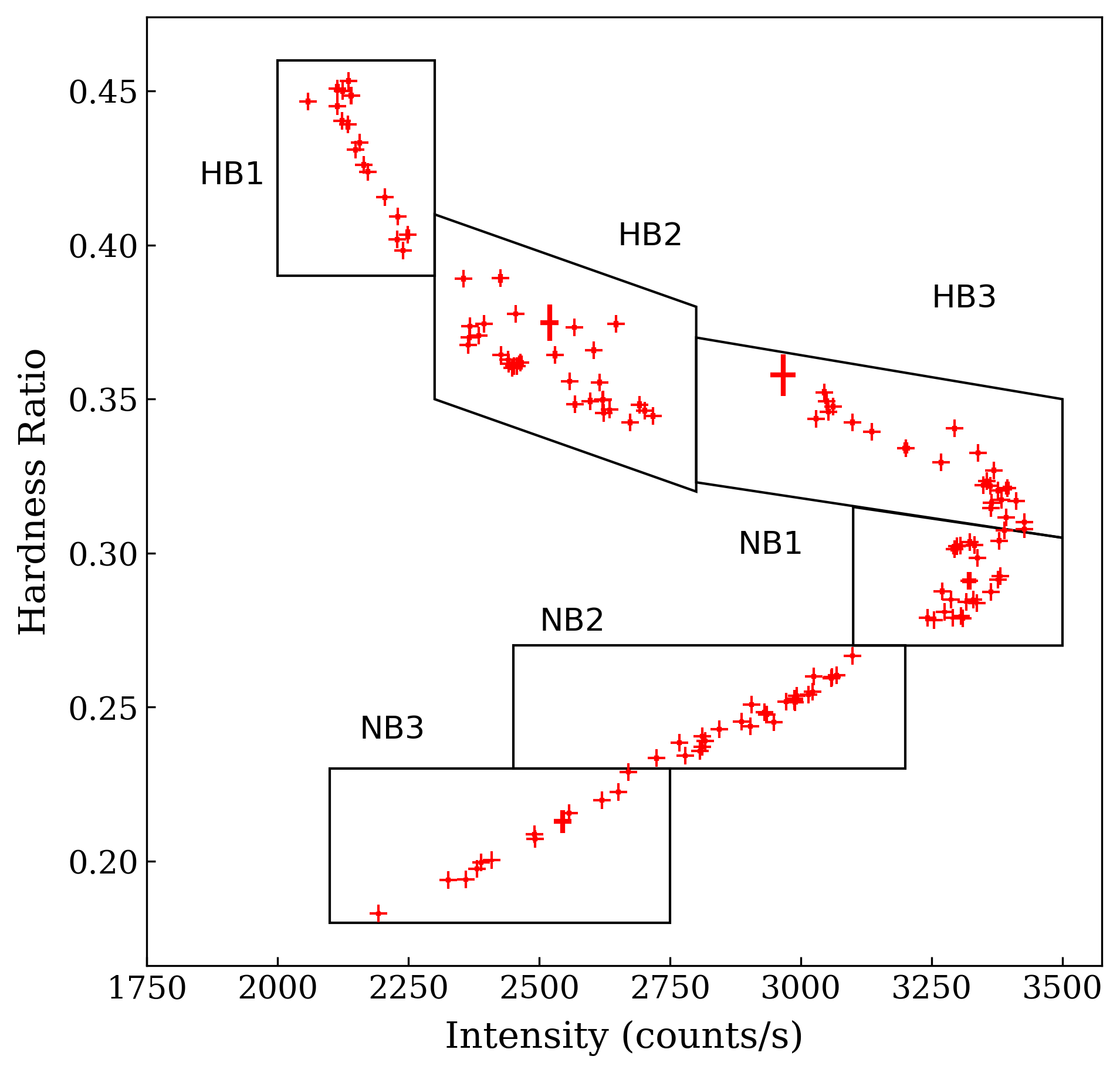}
    \caption{The hardness–intensity diagram (HID) of the source GX 5-1 plotted using data from LAXPC20 with hard color computed using the count rates in 8-20 keV divided by that in 3-8 keV and intensity represented by the total counts in 3-20 keV. Boxes represent the regions considered for segment-based spectral and timing analysis.}
    \label{HID_seg}
\end{figure}

We investigate the evolution of the source during the period of observation by plotting the Hardness Intensity Diagram (HID). Data from the LAXPC20 instrument was used for plotting the HID with 3–8 and 8–20 keV as the soft and hard energy bands, respectively, as shown in Figure \ref{HID_seg}. Each point in the figure corresponds to the intensity and hardness ratio values for the 256-s binned light curve. The source traces most of its "Z" track during these observations, except for the flaring branch. An extended horizontal branch (HB) is observed in the HID. For a detailed understanding of the evolution of spectral and temporal properties along the branches, we split the HID into six segments: three segments in HB and three in NB. To extract the segment-resolved spectral and light curve products, we identified the GTI files for each of the segments and used these files for product extraction. Spectral files and Power Density Spectra (PDS) are generated from each segment for further analysis.

\subsection{Timing Analysis} \label{sec:timing}
For timing analysis, only the data from the LAXPC20 instrument was considered. The evolution of QPO properties of the source along different branches was studied using PDS extracted in the energy range of 3–30 keV up to a maximum frequency of 100 Hz. Higher frequencies were excluded because the PDS above 100 Hz was dominated by Poisson noise. We calculated the rms normalized power density spectrum (\cite{Belloni_2002}) for each segment. Dead-time corrected Poisson noise level is subtracted from the PDS (\cite{Zang1995}, \cite{deadtime_Agrawal}). An instrumental peak around 50 Hz has been observed in the LAXPC20 instrument and is attributed to
the noise in the amplifier for anode A1 in LAXPC20 \cite{Antia_2017}. To account for this, we ignored frequencies close to 50 Hz from the PDS. The PDS was modeled using multiple Lorentzian and power law functions. QPOs and the band-limited noise (BLN) features in the PDS are modeled using the Lorentzian function (\cite{Belloni_2002}) defined as,
\begin{equation}
    A(E) = K(\sigma/2\pi)/[(\nu-\nu_{L})^{2} + (\sigma/2)^{2}]
\end{equation}
A Lorentzian model is characterized by three features: The centroid frequency ($\nu_{L}$) gives the position of the QPO signal in the PDS, the normalization ($K$), and the full width at half maximum (FWHM) of the signal ($\sigma$). The very low-frequency noise (VLFN) features in the PDS were modeled using a power-law function ($A\nu^{-\alpha}$) with normalization $A$ and index $\alpha$. The BLN features with an low value of $\nu_{L}$ are referred to as the low-frequency noise. We detect the presence of QPOs in five out of the six segments considered. The quality factor ($Q$), is defined as the ratio of centroid frequency to the FWHM of the QPO signal, and the root mean square amplitude, which is the measure of the strength of the QPO signal, was also calculated. 
We also report $\nu_{max}$, the frequency highest frequency covered by the Lorentzian component (\cite{Belloni_2002}).
\begin{equation}
    \nu_{max} = \sqrt{\nu_{L}^{2}+(\sigma/2)^{2}}
\end{equation}

This is the frequency at which the Lorentzian contributes most of its power per logarithmic frequency interval. The best-fit parameters of power spectral components in the different segments of the Z-track are reported in Table \ref{Timing}. PDS were generated in soft (3–4.5 keV) and hard (4.5-20.0 keV) energy bands to understand the dependence of the observed QPO features on energy. We find that the QPOs are observed only in PDS generated for the hard energy band, giving a strong connection between the origin of the QPOs and the high energy emission mechanism.

\subsection{Spectral Analysis} \label{sec:spectrum}

\begin{figure*}
	\centering
	\includegraphics[scale=0.48]{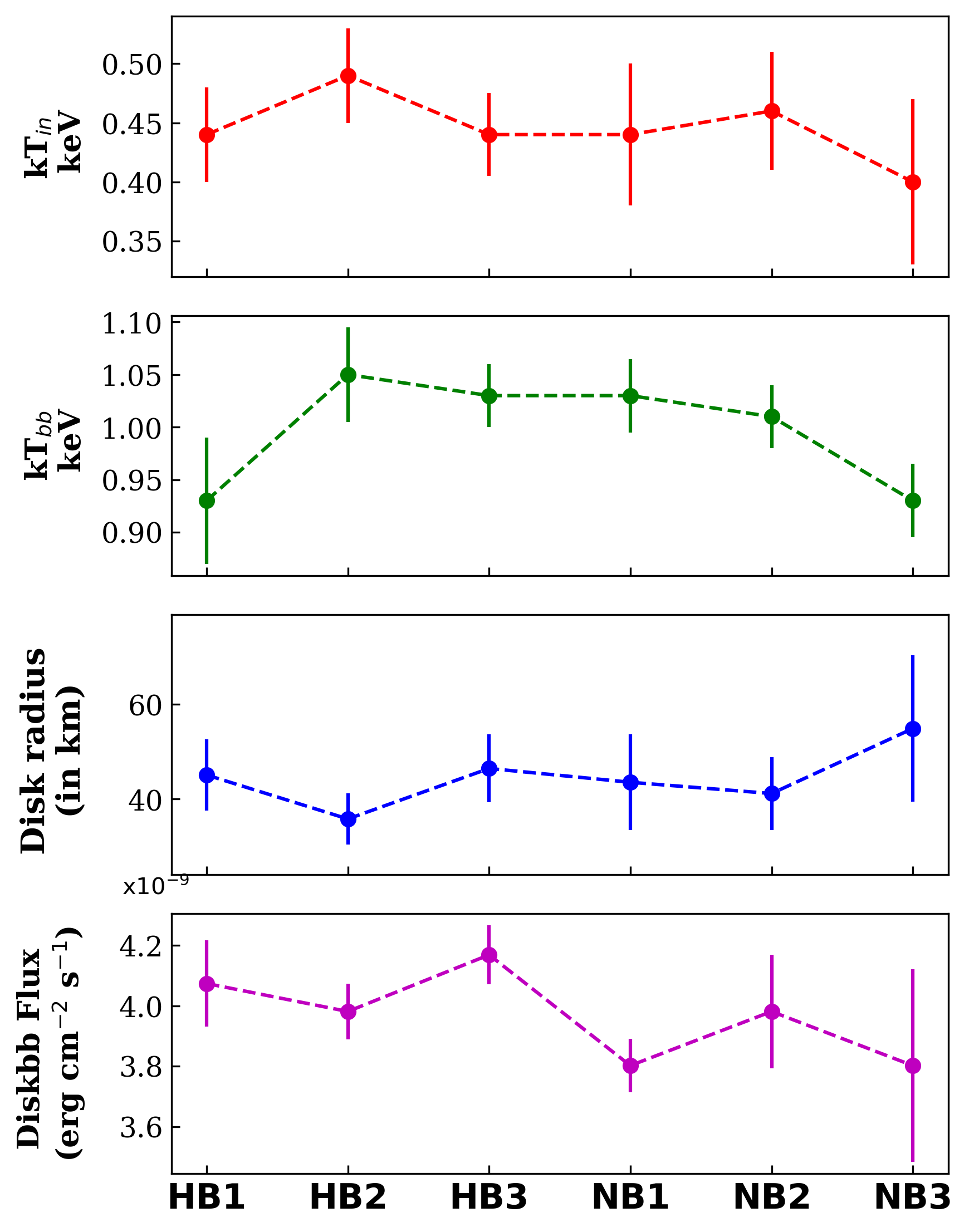}
	\includegraphics[scale=0.48]{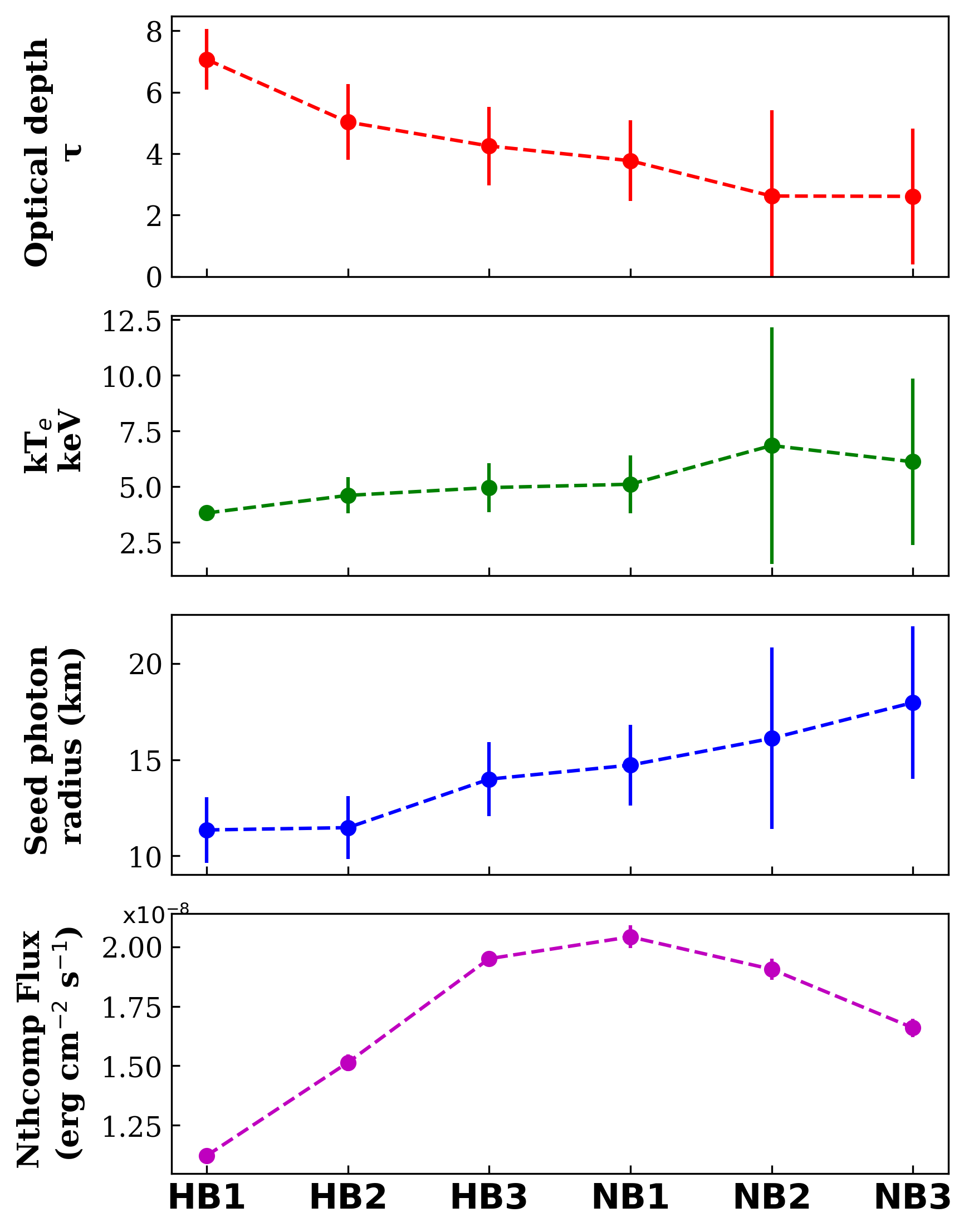}
	\caption{The evolution of spectral parameters from the \texttt{XSPEC} model \texttt{tbabs(diskbb+nthComp)} for the SXT and LAXPC combined spectral modeling of GX 5-1. The flux depicted in figure is the unabsorbed flux calculated using the \texttt{cflux} model in \texttt{XSPEC} in the energy range 0.3-50 keV. }
	\label{params_evolution}
\end{figure*}

Spectral fitting is carried out in the broad energy range of 0.7 to 20.0 keV by combining the data from LAXPC and SXT instruments. For SXT, 0.7–7.0 keV and 3.0–20.0 keV energy ranges for LAXPC are considered. We restricted the LAXPC energy range up to 20.0 keV due to the dominance of the background in the spectrum. We fitted the spectrum in six distinct segments in the HID using \texttt{XSPEC} version 12.0. We attempted two phenomenological spectral models to fit the X-ray spectrum of the source in all segments. First, we modeled the spectrum with the combination of a blackbody emission component from boundary-layer/NS-surface (\texttt{bbodyrad} model in \texttt{XSPEC}) and a thermal Comptonized emission from the corona (\texttt{nthComp}), which is called as Model 1. In Model 2, we replaced the blackbody emission with emission from a multi-color disk. The best-fit parameters are reported in Tables \ref{bbodyrad} and \ref{diskbb_fit}, respectively. Both models provide a statistically preferred fit; however, Model 2 was adopted as the best model because of the low reduced $\chi^{2}$ values observed in all segments. The spectra could be well described by the multi-color disc black body model \texttt{diskbb} (\cite{1984PASJ...36..741M}), along with the thermally Comptonized continuum emission model \texttt{nthComp} (\cite{1999MNRAS.309..561Z}). For absorption along the line of sight, the \texttt{XSPEC} model \texttt{Tbabs} (\cite{2000ApJ...542..914W}) was used. The hydrogen column density $N_{H}$ was fixed at the best-fit value of 2.7 $\times$ 10$^{22}$ cm$^{-2}$, and a systematic error of 3\% was added during the spectral fitting. The final model used for the fitting is \texttt{tbabs*(diskbb+nthComp)}. Spectral parameters are noted in each segment, and their evolution across different segments is plotted in Figure \ref{params_evolution}. The best-fit parameters are discussed in Table \ref{diskbb_fit} with 1$\sigma$ error of measurements. Reflection features such as Fe K$_{\alpha}$ or Compton hump were absent in the spectrum.

    \begin{table*}
	\renewcommand{\arraystretch}{1.3}
	\centering
        \addtocounter{table}{-1}
	\caption{The best-fit parameters obtained by fitting the PDS in different segments of the Z-track. The integrated rms in the 0.1–100 Hz is also provided.}
	\label{Timing}		
	\begin{tabular}{lcccccr} 
	\hline
        \hline
        Parameters & HB1 & HB2 & HB3 & NB1 & NB2 & NB3 \\
        \hline
        \multicolumn{7}{c}{QPO Lorentzian} \\
        \hline
        $\nu_{max}$ (Hz) & 12.3 $\pm$ 0.1 & 32.6 $\pm$ 0.7 & 40.3 $\pm$ 1.2 & 4.5 $\pm$ 0.3  & 5.8 $\pm$ 0.2 &  -- \\
        $\sigma$ (Hz) & 1.4 $\pm$ 0.2 & 5.1 $\pm$ 2.4 & 10.8 $\pm$ 3.5 & 1.6 $\pm$ 0.4 & 3.4 $\pm$ 0.7 & --  \\
        $LN$ ($\times 10^{-4}$) & 2.2 $\pm$ 0.3 & 5.4 $\pm$ 3.4 & 6.8 $\pm$ 1.3 & 41.5 $\pm$ 11.4 & 6.2 $\pm$ 0.8  &  -- \\
        $rms$ (\%) & 4.7 $\pm$ 0.3 & 2.3 $\pm$ 0.7 & 2.6 $\pm$ 0.3 & 6.4 $\pm$ 0.9 & 2.5 $\pm$ 0.2 &  -- \\
       
        \hline
        \multicolumn{7}{c}{Low frequency noise (Lorentzian)} \\
        \hline 
        $\nu_{max}$ (Hz) &  0$^*$ & 0$^*$ & 0$^*$ & 2.0 $\pm$ 0.3 & -- &  -- \\
        $\sigma$ (Hz) & 5.2 $\pm$ 0.4  & 6.1 $\pm$ 0.3 & 11 $\pm$ 3 & 1.7 $\pm$ 0.5 & -- &  -- \\
        $LN$ ($\times 10^{-3}$) &  6.7 $\pm$ 0.5 & 8.5 $\pm$ 0.3 & 1.7 $\pm$ 0.2 & 6.5 $\pm$ 1.7 & -- &  -- \\
        \hline
        \multicolumn{7}{c}{Power-law (VLFN)} \\
        \hline
        $\alpha$  & 1.2 $\pm$ 0.2  & 1.5 $\pm$ 1.0 & 1.1 $\pm$ 0.2 & 2.1 $\pm$ 1.0 & 1.7 $\pm$ 0.4 & 1.9 $\pm$ 0.5  \\
        $Norm$ ($\times 10^{-4}$) & 3.5 $\pm$ 1.1  & 0.8 $\pm$ 0.2 & 1.3 $\pm$ 0.5 & 3.0 $\pm$ 0.3 & 0.3 $\pm$ 0.1 &  0.4 $\pm$ 0.2 \\
        \hline
        \multicolumn{7}{c}{Additional Lorentzian 1} \\
        \hline
        $\nu_{max}$ (Hz) & 14.1 $\pm$ 0.1 & 26.3 $\pm$ 0.5 & -- & 14.2 $\pm$ 0.4 & -- &  -- \\
        $\sigma$ (Hz) & 1.3 $\pm$ 0.6 & 6.8 $\pm$ 0.6 & -- & 2.7 $\pm$ 0.7 & -- &  -- \\
        $LN$ ($\times 10^{-3}$) & 8.5 $\pm$ 4.3 & 3.1 $\pm$ 0.3 & -- & 0.6 $\pm$ 0.1 & -- & -- \\
        $rms$ (\%) & 2.9 $\pm$ 0.7 & 5.6 $\pm$ 0.2 & -- & 2.5 $\pm$ 0.2 & -- &  -- \\
        \hline
        \multicolumn{7}{c}{Additional Lorentzian 2} \\
        \hline
        $\nu_{max}$ (Hz) & 16.3 $\pm$ 0.5 & -- & -- & 24.1 $\pm$ 0.6 & -- &  -- \\
        $\sigma$ (Hz) & 8.7 $\pm$ 1.0  & -- & -- & 1.9 $\pm$ 1.5 & -- &  -- \\
        $LN$ ($\times 10^{-3}$) & 5.2 $\pm$ 0.7 & -- & -- & 0.1 $\pm$ 0.1 & -- & --  \\
        $rms$ (\%) & 7.2 $\pm$ 0.5 & -- & -- & 1.2 $\pm$ 0.3 & -- &  -- \\
        \hline
        \multicolumn{7}{c}{Additional Lorentzian 3} \\
        \hline
        $\nu_{max}$ (Hz) & -- & -- & -- & 48 $\pm$ 4 & -- &  -- \\
        $\sigma$ (Hz) & -- & -- & -- & 20$^*$ & -- &  -- \\
        $LN$ ($\times 10^{-4}$) & -- & -- & -- & 5.6 $\pm$ 1.2 & -- & --  \\
        $rms$ (\%) & -- & -- & -- & 2.4 $\pm$ 0.3 & -- &  -- \\
        $\chi^{2}/dof$ & 148/108 & 155/122 & 78/115 & 98/107 & 136/119 & 102/119 \\
        \hline
        \hline
        $^{*}$ represent frozen parameters
	\end{tabular}		
	\end{table*}

 \begin{figure*}
	\centering
	\includegraphics[width=0.95\textwidth]{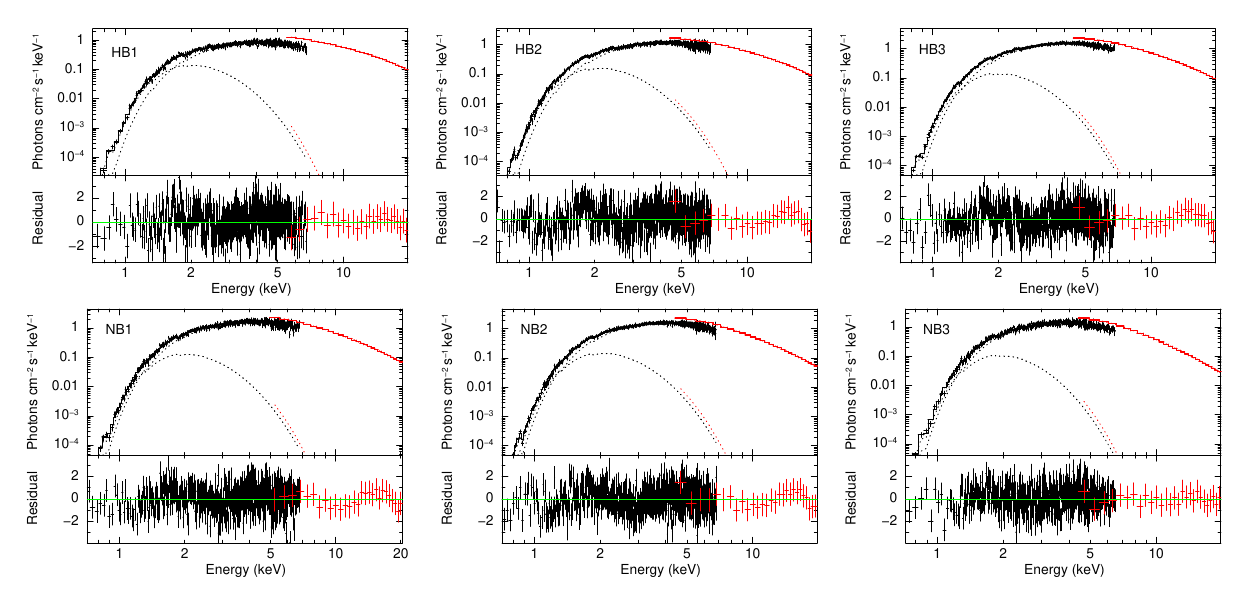}
	\caption{GX 5-1 spectrum in Horizontal and Normal branches fitted using the model \texttt{tbabs*(diskbb+nthComp)} in the 3-20 keV energy band. Residual are also plotted along with the fit.}
	\label{spec_pds}
	\end{figure*}

 	\begin{figure*}
	\centering
	\includegraphics[width=0.95\textwidth]{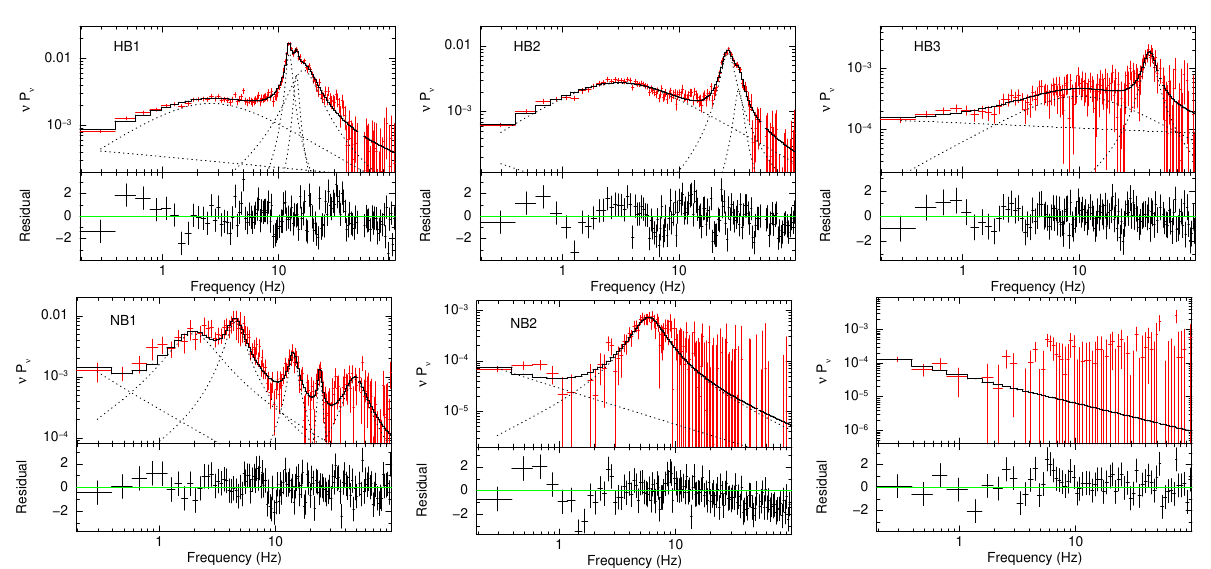}
	\caption{Power density spectrum in the Horizontal and Normal branch segments showing QPO features. The PDS is modeled using power-law and multiple Lorentzian functions.}
	\label{pds_fit}
	\end{figure*}
 
    \begin{table*}	
	\renewcommand{\arraystretch}{1.5}		
	\centering			
    \caption{Best-fit spectral parameter values along with their 1$\sigma$ error values for the individual segments in the HID for 3-20 keV energy band combining LAXPC and SXT spectrum. The model used for fitting is \textit{tbabs*(bbodyrad+nthComp)}.}
    \label{bbodyrad}  
	\begin{tabular}{lcccccr} 
	\hline
	\hline
	Fitted parameters & HB1 & HB2 & HB3 & NB1 & NB2 & NB3 \\
        \hline
        $kT$ (keV)  & 1.3$^{+0.1}_{-0.1}$ & 1.31$^{+0.05}_{-0.05}$ & 1.25$^{+0.03}_{-0.03}$ & 1.15$^{+0.05}_{-0.05}$ & 1.16$^{+0.03}_{-0.03}$ & 1.11$^{+0.03}_{-0.03}$ \\   
        
        $N_{bbodyrad}$ ($\times 10^{2}$) & 1.2$^{+0.3}_{-0.3}$ & 2.1$^{+0.4}_{-0.4}$ & 4.2$^{+0.5}_{-0.4}$ & 4.8$^{+1.1}_{-0.9}$ & 9.1$^{+1.0}_{-1.1}$ & 9.4$^{+1.1}_{-1.3}$ \\   
        
        $\Gamma$  & 1.77$^{+0.04}_{-0.06}$ & 1.8$^{+0.1}_{-0.1}$ & 1.7$^{+0.1}_{-0.1}$ & 2.2$^{+0.1}_{-0.2}$ & 1.8$^{+0.1}_{-0.1}$ & 2.1$^{+0.1}_{-0.1}$ \\    
        
        $kT_{e}$ (keV)  & 3.2$^{+0.2}_{-0.2}$ & 3.2$^{+0.1}_{-0.2}$ & 3.0$^{+0.1}_{-0.1}$ & 3.2$^{+0.1}_{-0.1}$ & 2.8$^{+0.1}_{-0.1}$ & 2.8$^{+0.2}_{-0.1}$ \\        
        
        $kT_{bb}$ (keV)  & 0.4$^{+0.1}_{-0.1}$ & 0.6$^{+0.1}_{-0.2}$ & 0.5$^{+0.1}_{-0.2}$ & 1.4$^{+0.2}_{-0.3}$ & 0.6$^{+0.2}_{-0.3}$ & 0.84$^{*}$ \\       
        
        $N_{nth}$  & 1.3$^{+0.1}_{-0.1}$ & 1.25$^{+0.1}_{-0.1}$ & 1.3$^{+0.1}_{-0.1}$ & 1.1$^{+0.2}_{-0.2}$ & 1.2$^{+0.2}_{-0.3}$ & 1.92$^{+0.2}_{-0.2}$  \\ 
        
        $\chi^{2}/dof$ & 871/645  & 969/649  &  890/651 &  850/651  & 859/651 & 772/629\\
        \hline
        \hline
        $^{*}$ represent frozen parameters
    \end{tabular}
    \end{table*}

    \begin{table*}		
	\renewcommand{\arraystretch}{1.5}		
	\centering		
    \caption{Best-fit spectral parameter values along with their 1$\sigma$ error values for the individual segments in the HID for 3-20 keV energy band combining LAXPC and SXT spectrum. The model used for fitting is \textit{tbabs*(diskbb+nthComp)}.}
    \label{diskbb_fit}	
	\begin{tabular}{lcccccr} 
	\hline
	    \hline
         Fitted parameters& HB1 & HB2 & HB3 & NB1 & NB2 & NB3 \\
         \hline
         $kT_{in}$ (keV)  & 0.44$^{+0.04}_{-0.04}$ & 0.49$^{+0.04}_{-0.04}$ & 0.44$^{+0.03}_{-0.04}$ & 0.44$^{+0.05}_{-0.07}$ & 0.46$^{+0.05}_{-0.05}$ & 0.40$^{+0.06}_{-0.08}$  \\
         
         $N_{diskbb}$ ($\times 10^{3}$)  & 5.8$^{+1.9}_{-3.1}$ & 3.6$^{+1.1}_{-1.6}$ & 6.1$^{+1.9}_{-2.8}$ & 5.4$^{+2.5}_{-4.7}$ & 4.8$^{+1.8}_{-2.9}$ & 8.5$^{+4.8}_{-11}$ \\
         
         $\Gamma$  & 2.3$^{+0.1}_{-0.1}$ & 2.7$^{+0.2}_{-0.2}$ & 3.0$^{+0.2}_{-0.2}$ & 3.2$^{+0.2}_{-0.3}$ & 3.7$^{+0.3}_{-0.4}$ & 3.9$^{+0.3}_{-0.4}$ \\
         
         $kT_{e}$ (keV)  & 3.8$^{+0.3}_{-0.4}$ & 4.6$^{+0.6}_{-1.0}$ & 5.0$^{+0.7}_{-1.5}$ & 5.1$^{+0.8}_{-1.8}$ & 6.8$^{+1.9}_{-8.8}$ & 6.1$^{+1.5}_{-5.9}$  \\
         
         $kT_{bb}$ (keV)  & 0.93$^{+0.06}_{-0.06}$ & 1.05$^{+0.04}_{-0.05}$ & 1.03$^{+0.03}_{-0.03}$ & 1.03$^{+0.03}_{-0.04}$ & 1.01$^{+0.03}_{-0.03}$ & 0.93$^{+0.03}_{-0.04}$  \\
         
         $N_{nth}$  & 0.40$^{+0.04}_{-0.03}$ & 0.49$^{+0.04}_{-0.04}$ & 0.69$^{+0.03}_{-0.03}$ & 0.75$^{+0.05}_{-0.05}$ & 0.78$^{+0.05}_{-0.04}$ & 0.82$^{+0.06}_{-0.06}$ \\

         flux$_{\text{disc}}$\footnote{All flux calculated in the 0.3-50 keV energy band}($\times 10^{-9}$)  & 4.1 $\pm$ 0.1 & 4.0 $\pm$ 0.1 & 4.2 $\pm$ 0.1 & 3.8 $\pm$ 0.1 & 4.0 $\pm$ 0.2 & 3.8 $\pm$ 0.3  \\
         
         flux$_{comp}$ ($\times 10^{-8}$)  & 1.12 $\pm$ 0.01 & 1.51 $\pm$ 0.02 & 1.94 $\pm$ 0.02 & 2.04 $\pm$ 0.02 & 1.90 $\pm$ 0.02 & 1.65 $\pm$ 0.02  \\ 
         
         Total flux ($\times 10^{-8}$) & 2.8 $\pm$ 0.4 & 3.0 $\pm$ 0.4 & 3.9 $\pm$ 0.5 & 3.7 $\pm$ 0.4 & 3.5 $\pm$ 0.7 & 3.2 $\pm$ 0.8 \\
         
         $\chi^{2}$/dof & 689/649 & 682/629 & 691/629 & 687/629 & 664/627 & 577/598 \\
         
         \hline
         Derived parameters &&&&&& \\
         Disc Radius (km)  & 45 $\pm$ 8 & 36 $\pm$ 5 & 46 $\pm$ 7 & 44 $\pm$ 10 & 41 $\pm$ 8 & 55 $\pm$ 16  \\
         
         $y$ parameter & 1.5 $\pm$ 0.4 & 0.9 $\pm$ 0.4 & 0.7 $\pm$ 0.4 & 0.6 $\pm$ 0.4 & 0.4 $\pm$ 0.8 & 0.3 $\pm$ 0.6 \\
         
         Optical depth ($\tau$) & 7.1 $\pm$ 1.0 & 5.0 $\pm$ 1.2 & 4.2 $\pm$ 1.3 & 3.8 $\pm$ 1.3 & 2.6 $\pm$ 2.8 & 2.6 $\pm$ 2.2  \\
         
         Seed photon radius (km) & 11.4 $\pm$ 1.7 & 11.5 $\pm$ 1.6 & 14.0 $\pm$ 1.9 & 14.7 $\pm$ 2.1 & 16.1 $\pm$ 4.7 & 18.0 $\pm$ 4.0 \\
         \hline
         \hline
    \end{tabular}
    \end{table*}


\section{RESULTS} \label{sec: results and discussion}

The broad-band spectral analysis of GX 5-1 has been carried out using the \textit{\textit{AstroSat}} SXT and LAXPC20 instruments. The temporal properties of the source have been studied using data from the LAXPC20 instrument. The HID of the source traces the HB and NB branches of the Z pattern. The broad-band spectrum in the energy range 0.7–20.0 keV was used to understand the spectral properties and estimate various physical parameters of the source. The simultaneous SXT and LAXPC spectrum could be modeled using the \texttt{tbabs*(diskbb+nthComp)} model in \texttt{XSPEC}. The best-fit values of the spectral parameters are given in Table \ref{diskbb_fit} and spectral fit in all segments considered are shown in Figure \ref{spec_pds}. In this section, we discuss the findings and understanding of the source behavior.

\subsection{Estimation of physical parameters}

We derive key parameters, such as inner disc radius and optical depth, for the spectrum in each segment from spectral fitting. The normalization of the black body component of the disc gives us an estimate of the inner radius of the accretion disc. It is given by the relation,

\begin{equation}
	Norm = \Bigg(\frac{R_{in}}{D_{10}}\Bigg)^{2} \cos\theta
\end{equation}
where $D_{10}$ is the distance to the source in units of 10 kpc and $\theta$ the inclination of the disc. The Monte Carlo simulation of the X-ray halos of GX 5-1 suggests that the approximate distance to the source is 4.2 kpc (\cite{2018ApJ...852..121C}).We see that the radius is approximately in the range of 35–55 km in all the segments and shows an increase along the HB.
To calculate the disk radius, we assume an inclination of  60$^{\circ}$ for the source. This assumption is based on the absence of dips or eclipses in its lightcurve, as noted by \cite{Christian_1997}. The sources that do exhibit dips typically have an inclination greater than 70$^{\circ}$ .Similar assumptions have been considered in the past for this source (\cite{Bhulla_2019}, \cite{sriram_2022}). 

The optical depth is found from the photon index, assuming a spherical geometry and uniform density for the corona. It can be found out from spectral fitting using the below equation \cite{1996A&AS..120C.553Z},
\begin{equation}	
	\Gamma = - \frac{1}{2} + \sqrt{\frac{9}{4}+\frac{1}{\frac{kT_{e}}{mc^{2}}(1+\frac{\tau}{3})\tau}}	
\end{equation}
where $\Gamma$ is the photon index and $kT_{e}$ is the electron temperature in the plasma. We estimate $\tau$ for each segment and see a decreasing trend as the source moves along HB, as shown in Figure \ref{params_evolution}.

The \texttt{cflux} model was used to calculate the unabsorbed flux for both the disc and Comptonization component in the energy range 0.3 to 50.0 keV, and the variation in flux for both components as the source moves along the Z track is plotted in Figure \ref{params_evolution}. 

We calculated the equivalent seed photon emission area, which is done by equalizing the bolometric luminosity of the soft photons to that of a black body with a temperature of $kT_{bb}$ and a radius of $R_{bb}$  (\cite{Rw}).

\begin{equation}
    R_{bb} = 3\times 10^{4} \sqrt{\frac{f_{bol}}{1+y}} / (kT_{bb})^{2}
\end{equation}

\noindent
Where $y$ is the Comptonization parameter given by $y = 4kT_{e}/m_{e}c^{2}$, $f_{bol}$ is the bolometric luminosity of the source in the unit erg/m$^{2}$/s, and $d$ is the distance to the source in kpc. The average value of $R_{W}$ thus obtained is $\sim$14 km, and it also shows an increasing trend from HB to NB.

\subsection{Spectral behaviour of the source}
The broadband spectrum of the source in the energy range of 0.7–20 keV was modeled using a disc blackbody and a thermal Comptonization model. The unabsorbed flux of the Comptonization and disc blackbody components was computed using the \texttt{cflux} command for all segments in the HID. At the beginning of the observation, GX 5-1 was in the HB. As the source goes to the Hard Apex (HA), the Comptonization flux increases along the HB. The flux then decreases as it moves through the NB. To understand the spectral evolution of the source along the Z-track, we plot the inner disc temperature ($kT_{in}$), electron temperature ($kT_{e}$), the inner radius of the disc, and optical depth as a function of different positions in the HID. These parameters were derived from the spectral fitting done in all segments of HID. The variation of the parameters is shown in Figure \ref{params_evolution}. 
The electron temperature $kT_{e}$ of the corona and the $\Gamma$ increases along HB. $kT_{e}$ increases from $\sim$3.82 keV in HB1 to $\sim$6.12 keV in NB3. During this phase of the evolution from HB to HA, we see that the optical depth of the corona ($\tau$) decreases. The evolution of the source from HA to NB is associated with a decrease in the blackbody temperature ($kT_{bb}$). No clear trend in the variation of inner disc radius was seen during the evolution of the source.

\subsection{Temporal behaviour of the source}

We investigate the presence of QPOs in the PDS generated for all segments in the HID. We found the presence of HBOs in all segments of the HB (HB1, HB2, and HB3) and NBOs in two segments (upper and middle NB). No oscillation feature is found in the lower NB. In the extended section of the HB (HB1), we observe HBO at $\sim$14.12 Hz, and the frequency increases to $\sim$40 Hz as the source moves along the horizontal branch in the Z-track. Normal branch oscillations (NBOs) are detected at $\sim$5 Hz in two segments, with QPO frequency showing an increasing trend towards the soft apex. PDS generated in all the HID segments are modeled using multiple Lorentzian functions as shown in Figure \ref{pds_fit}, the frequency, width, and amplitude of the QPO signal along with the rms amplitude and maximum frequency are reported in Table \ref{Timing}. The QPO seen in the segments HB1 and HB2 had a shoulder-like component on the high-frequency side. The complete feature was modeled using a combination of two Lorentzian. In NB1, along with the QPO at 4.45 Hz, additional peak features at $\sim$14 Hz, $\sim$24 Hz, and a broad feature at $\sim$48 Hz are observed in the PDS.

\section{Discussion} \label{sec:summary}
This work reports the results obtained from a detailed spectral and timing analysis of the neutron star LMXB source GX 5-1 using \textit{AstroSat} observations. We investigate the spectral and temporal evolution of the source as it traces the Z pattern in the HID. During the observations, the source traced a complete HB and NB. The broadband spectra in the range of 0.7–20 keV could be well modeled using a combination of a multicolor disk blackbody model and a Comptonization model. We observed the presence of QPOs in HB and NB of the Z-track with increasing QPO frequency in HB towards the hard apex. The variation of the spectral parameters and their association with the QPO properties are also studied.

\subsection{Evolution of spectral properties}

We attempt to understand the origin of Z-track evolution in the source GX 5-1 using spectral analysis done with the simultaneous \textit{AstroSat}/SXT and LAXPC observations. We considered two different approaches to study the spectral behavior of the source by modeling the source spectrum in the 0.7-20.0 keV. Out of the two models, the source spectrum could be well fitted using a combination of a multi-color disc blackbody model to describe the emission from the accretion disc and a thermal Comptonization model for emission from the corona. 

The inner disc temperature remains almost constant throughout the Z-track varying from 0.40 to 0.49 keV. No systematic variation or clear trend is observed in the case of disc radius, which has an average value of $\sim$45 km. However, we observe that an increase in the disc inner radius corresponds to a decrease in the inner disc temperature, which is expected. We see that, along the HB, disc parameters remain almost constant as the source evolves, considering the temperature, disc inner radius, and disc flux. However, certain parameters related to Comptonization such as the Comptonization flux, exhibit an increasing trend, while the optical depth of the corona shows a decreasing trend, as illustrated in Figure \ref{params_evolution}. The Comptonization flux increases along the HB, reaching its maximum value before it starts to decrease along the NB. In the HB, it shows a positive correlation with the electron temperature of the corona. 
The optical depth ($\tau$) of the corona is found to decrease along the HB and along the NB ($\sim$7.07 in HB1 to $\sim$2.61 in NB3) which is consistent with the behavior exhibited by GX 17+2 (\cite{2020Ap&SS.365...41A}). This behavior of the source could be described in terms of an increasing accretion rate scenario as presented in \cite{2003MNRAS.346..933A}. An increase in the soft photon supply to the corona results in the cooling and settling down of a part of the material from the corona onto the disc. This can result in the reduction of the optical depth of the corona. The seed photon radius calculated from the best-fit values shows an increasing trend as the source evolves from HB to NB from $\sim$11.35 km in HB1 to $\sim$17.97 in NB3. The average value is 14 km, which is much shorter than the inner disc radius observed in any segment of the HID. This indicates that the seed photons are emitted from a region much closer to the neutron star's surface. 

The absence of systematic variation observed in the disk parameters suggests that the source movement along the Z-track could be driven by the variation in the corona or boundary layer and not primarily by the change in mass accretion rate. This result is in agreement with that of \cite{Homan_2002} and \cite{Lin_2012}. The optical depth decreases along the horizontal branch, which is associated with an increase in the Comptonization flux. However, the disc flux remains constant. As the source moves towards the NB, some portion of the corona cools and settles down onto the disc, causing the optical depth to decrease. At the same time, the other portion of the corona is hotter than before, causing a rise in the coronal temperature. In the Normal Branch, the optical depth continues to decrease, followed by a decrease in the  Comptonization flux, which drives the source evolution along NB.

\subsection{Evolution of timing properties}

All the segments in the HID, except NB3, show QPOs in the PDS. We observe HBO with frequencies in the range of 13-40 Hz. \cite{Bhulla} have reported QPOs at $\sim$50 Hz in this source using AstroSat data. These observed frequencies are in the expected frequency ranges for Z sources (\cite{10.1093/mnras/231.2.379}). The HBO frequency increases as the source moves along the HB towards the hard apex. This frequency change in HB is associated with a decrease in the strength of the HBO. The HBOs and NBOs are observed only in the hard energy ($>$ 5 keV) band, which is consistent with previous studies showing that the fractional rms increases with energy (\cite{10.1093/mnras/231.2.379}, \cite{Jonker_2000}). This indicates that the HBOs and NBOs have a stronger contribution from Comptonizing corona. This behavior is similar to what was observed for the Z source GX 334+0 \cite{Bhargava_2023}.

We attempt to explain the observed QPO frequencies using the Lense-Thirring precession model(\cite{10.1111/j.1745-3933.2009.00693.x}), where the observed low-frequency (LF) QPOs in black hole and neutron star systems are attributed to the misalignment of the inner hot accretion flow with the compact object spin. The frequency of the LF QPOs is given by equation (2) from \cite{10.1111/j.1745-3933.2009.00693.x}:
\begin{equation}
 v_{\text {prec }}=\frac{(5-2 \zeta) a\left(1-\left(r_i / r_o\right)^{1 / 2+\zeta}\right)}{\pi(1+2 \zeta) r_o^{5 / 2-\zeta} r_i^{1 / 2+\zeta}\left[1-\left(r_i / r_o\right)^{5 / 2-\zeta}\right]} \frac{c}{R_g} 
\end{equation}
Where the mass distribution of the hot flow near the compact object is parameterized by $\zeta$, $r_i$ and $r_o$ are the inner and outer radii of the hot flow, respectively, and $a$ is the spin parameter. According to the simulation performed by \cite{Fragile_2007}, $\zeta$ can be equated to zero for black holes and can also be used for Neutron star sources. This scenario may lead to an increased concentration of the accretion flow on the neutron star's surface as the accretion rate increases. To compute the precision frequency, the outer radius of the inner hot flow was kept at the inner disc radius $R_{in}$ obtained from spectral analysis. The outer radius was calculated for the observed QPO frequencies. The model could predict the QPO frequencies observed in HB (13.56, 26.49 Hz); however, it failed to predict any QPOs above 26 Hz using the given parameters due to the larger inner disc. The model could not explain the observed QPO in the HB3 with a centroid frequency of 40 Hz.

We observe NBOs with a frequency $\sim$5 Hz in two of the segments considered for analysis. In NB1, along with NBO at $\sim$4 Hz, we observe other QPO features at $\sim$14 Hz and $\sim$24 Hz and a broad feature at $\sim$48 Hz. The origin of these features is not understood. A model for NBO was proposed by \cite{Titarchuk_2001}, in which the NBOs are considered as viscous oscillations of a spherical shell around the neutron star surface. The spherical shell viscous frequency $\nu_{ssv}$ is given by,
\begin{equation}
\nu_{ssv} = \frac{f\nu_{s}}{L}
\end{equation}
Where $L$ is the size of the viscous shell, and $f$ can be 0.5 and $1/(2\pi)$ for the stiff and free boundary conditions in the transition layer, respectively. $\nu_{s}$ is the sonic velocity that is estimated based on Equation (3) from \cite{Hasinger_1987}.

\begin{equation}
    \nu_{s} = 4.2 \times 10^{7} R_{6}^{-1/4}\Bigg(\frac{M}{M_{\odot}}\frac{L}{L_{edd}}\Bigg)^{1/8} \text{cm s$^{-1}$}
\end{equation}

Where $R_{6}$ is the radius of the neutron star in units of $10^{6}$ cm. For calculations, we consider a radius of 10 km and a mass of 1.4 $M_{\odot}$ for the neutron star. We estimate $L_{s}$ to be 32 km for an NBO frequency of 6 Hz. This value is found to be greater than the calculated seed photon radius in NB. The inner region between the accretion disk, with an average inner radius of 45 km, and the NS surface could be hosting a boundary layer or a transition shell, whose oscillations can produce the NBO.

\vspace{0.1cm}
\section*{ACKNOWLEDGMENTS} \label{Acknowledgement}

This work uses data from the \textit{AstroSat} mission of ISRO archived at Indian Space Science Data Centre (ISSDC). The article has used data from the SXT and the LAXPC developed at TIFR, Mumbai, and the \textit{AstroSat} POCs at TIFR are thanked for verifying and releasing the data via the ISSDC and providing the necessary software tools. Authors thank GD, SAG; DD, PDMSA and Director, URSC for encouragement and support to carry out this research. We have used data and/or software provided by the High Energy Astrophysics Science Archive Research Centre (HEASARC), which is a service of the Astrophysics Science Division at NASA/GSFC and the High Energy Astrophysics Division of the Smithsonian Astrophysical Observatory.

\bibliography{GX5-1_Astrosat}{}

\begin{thebibliography}{}
\expandafter\ifx\csname natexlab\endcsname\relax\def\natexlab#1{#1}\fi
\providecommand{\url}[1]{\href{#1}{#1}}
\providecommand{\dodoi}[1]{doi:~\href{http://doi.org/#1}{\nolinkurl{#1}}}
\providecommand{\doeprint}[1]{\href{http://ascl.net/#1}{\nolinkurl{http://ascl.net/#1}}}
\providecommand{\doarXiv}[1]{\href{https://arxiv.org/abs/#1}{\nolinkurl{https://arxiv.org/abs/#1}}}

\bibitem[{Agrawal {et~al.}(2018)Agrawal, Nandi, Girish, \& Ramadevi}]{deadtime_Agrawal}
Agrawal, V.~K., Nandi, A., Girish, V., \& Ramadevi, M.~C. 2018, Monthly Notices of the Royal Astronomical Society, 477, 5437, \dodoi{10.1093/mnras/sty1005}

\bibitem[{{Agrawal} {et~al.}(2020){Agrawal}, {Nandi}, \& {Ramadevi}}]{2020Ap&SS.365...41A}
{Agrawal}, V.~K., {Nandi}, A., \& {Ramadevi}, M.~C. 2020, \apss, 365, 41, \dodoi{10.1007/s10509-020-3748-0}

\bibitem[{{Agrawal} \& {Sreekumar}(2003)}]{2003MNRAS.346..933A}
{Agrawal}, V.~K., \& {Sreekumar}, P. 2003, \mnras, 346, 933, \dodoi{10.1111/j.1365-2966.2003.07147.x}

\bibitem[{{Antia} {et~al.}(2017){Antia}, {Yadav}, {Agrawal}, {Verdhan Chauhan}, {Manchanda}, {Chitnis}, {Paul}, {Dedhia}, {Shah}, {Gujar}, {Katoch}, {Kurhade}, {Madhwani}, {Manojkumar}, {Nikam}, {Pandya}, {Parmar}, {Pawar}, {Pahari}, {Misra}, {Navalgund}, {Pandiyan}, {Sharma}, \& {Subbarao}}]{Antia_et_al}
{Antia}, H.~M., {Yadav}, J.~S., {Agrawal}, P.~C., {et~al.} 2017, \apjs, 231, 10, \dodoi{10.3847/1538-4365/aa7a0e}

\bibitem[{Antia {et~al.}(2017)Antia, Yadav, Agrawal, Chauhan, Manchanda, Chitnis, Paul, Dedhia, Shah, Gujar, Katoch, Kurhade, Madhwani, Manojkumar, Nikam, Pandya, Parmar, Pawar, Pahari, Misra, Navalgund, Pandiyan, Sharma, \& Subbarao}]{Antia_2017}
Antia, H.~M., Yadav, J.~S., Agrawal, P.~C., {et~al.} 2017, The Astrophysical Journal Supplement Series, 231, 10, \dodoi{10.3847/1538-4365/aa7a0e}

\bibitem[{Asai {et~al.}(2000)Asai, Dotani, Nagase, \& Mitsuda}]{Asai_2000}
Asai, K., Dotani, T., Nagase, F., \& Mitsuda, K. 2000, The Astrophysical Journal Supplement Series, 131, 571, \dodoi{10.1086/317374}

\bibitem[{Belloni {et~al.}(2002)Belloni, Psaltis, \& van~der Klis}]{Belloni_2002}
Belloni, T., Psaltis, D., \& van~der Klis, M. 2002, The Astrophysical Journal, 572, 392–406, \dodoi{10.1086/340290}

\bibitem[{Bhargava {et~al.}(2023)Bhargava, Bhattacharyya, Homan, \& Pahari}]{Bhargava_2023}
Bhargava, Y., Bhattacharyya, S., Homan, J., \& Pahari, M. 2023, The Astrophysical Journal, 955, 102, \dodoi{10.3847/1538-4357/acee7a}

\bibitem[{Bhulla {et~al.}(2019)Bhulla, Misra, Yadav, \& A~Jaaffrey}]{Bhulla_2019}
Bhulla, Y., Misra, R., Yadav, J.~S., \& A~Jaaffrey, S.~N. 2019, Research in Astronomy and Astrophysics, 19, 114, \dodoi{10.1088/1674-4527/19/8/114}

\bibitem[{{Bhulla} {et~al.}(2019){Bhulla}, {Misra}, {Yadav}, \& {Jaaffrey}}]{Bhulla}
{Bhulla}, Y., {Misra}, R., {Yadav}, J.~S., \& {Jaaffrey}, S.~N.~A. 2019, Research in Astronomy and Astrophysics, 19, 114, \dodoi{10.1088/1674-4527/19/8/114}

\bibitem[{{Bradt} {et~al.}(1968){Bradt}, {Naranan}, {Rappaport}, \& {Spada}}]{1968ApJ...152.1005B}
{Bradt}, H., {Naranan}, S., {Rappaport}, S., \& {Spada}, G. 1968, \apj, 152, 1005, \dodoi{10.1086/149613}

\bibitem[{Capitanio {et~al.}(2023)Capitanio, Gnarini, Fabiani, Ursini, Farinelli, Cocchi, Cavero, \& Marra}]{polarised_XRBs}
Capitanio, F., Gnarini, A., Fabiani, S., {et~al.} 2023, Polarised light from accreting low mass X-ray binaries.
\newblock \doarXiv{2312.14779}

\bibitem[{{Chiranjeevi} \& {Sriram}(2022)}]{2022MNRAS.516.2500C}
{Chiranjeevi}, P., \& {Sriram}, K. 2022, \mnras, 516, 2500, \dodoi{10.1093/mnras/stac2319}

\bibitem[{Christian \& Swank(1997)}]{Christian_1997}
Christian, D.~J., \& Swank, J.~H. 1997, The Astrophysical Journal Supplement Series, 109, 177, \dodoi{10.1086/312970}

\bibitem[{{Clark}(2018)}]{2018ApJ...852..121C}
{Clark}, G.~W. 2018, \apj, 852, 121, \dodoi{10.3847/1538-4357/aaa1f0}

\bibitem[{Fabiani {et~al.}(2023)Fabiani, Capitanio, Iaria, Poutanen, Gnarini, Ursini, Farinelli, Bobrikova, Steiner, Svoboda, Anitra, Baglio, Carotenuto, Santo, Ferrigno, Lewis, Russell, Russell, van~den Eijnden, Cocchi, Marco, Monaca, Liu, Rankin, Weisskopf, Xie, Bianchi, Burderi, Salvo, Egron, Illiano, Kaaret, Matt, Mikušincová, Muleri, Papitto, Agudo, Antonelli, Bachetti, Baldini, Baumgartner, Bellazzini, Bongiorno, Bonino, Brez, Bucciantini, Castellano, Cavazzuti, Chen, Ciprini, Costa, Rosa, Monte, Gesu, Lalla, Donnarumma, Doroshenko, Dovčiak, Ehlert, Enoto, Evangelista, Ferrazzoli, Garcia, Gunji, Hayashida, Heyl, Iwakiri, Jorstad, Karas, Kislat, Kitaguchi, Kolodziejczak, Krawczynski, Latronico, Liodakis, Maldera, Manfreda, Marin, Marinucci, Marscher, Marshall, Massaro, Mitsuishi, Mizuno, Negro, Ng, O'Dell, Omodei, Oppedisano, Pavlov, Peirson, Perri, Pesce-Rollins, Petrucci, Pilia, Possenti, Puccetti, Ramsey, Ratheesh, Roberts, Romani, Sgrò, Slane, Soffitta, Spandre, Swartz, Tamagawa, Tavecchio,
  Taverna, Tawara, Tennant, Thomas, Tombesi, Trois, Tsygankov, Turolla, Vink, Wu, \& Zane}]{fabiani2023discovery}
Fabiani, S., Capitanio, F., Iaria, R., {et~al.} 2023, Discovery of a variable energy-dependent X-ray polarization in the accreting neutron star GX 5-1.
\newblock \doarXiv{2310.06788}

\bibitem[{{Fiocchi} {et~al.}(2007){Fiocchi}, {Bazzano}, {Ubertini}, \& {Zdziarski}}]{2007ApJ...657..448F}
{Fiocchi}, M., {Bazzano}, A., {Ubertini}, P., \& {Zdziarski}, A.~A. 2007, \apj, 657, 448, \dodoi{10.1086/510573}

\bibitem[{{Fisher} {et~al.}(1968){Fisher}, {Jordan}, {Meyerott}, {Acton}, \& {Roethig}}]{1968ApJ...151....1F}
{Fisher}, P.~C., {Jordan}, W.~C., {Meyerott}, A.~J., {Acton}, L.~W., \& {Roethig}, D.~T. 1968, \apj, 151, 1, \dodoi{10.1086/149414}

\bibitem[{Fragile {et~al.}(2007)Fragile, Blaes, Anninos, \& Salmonson}]{Fragile_2007}
Fragile, P.~C., Blaes, O.~M., Anninos, P., \& Salmonson, J.~D. 2007, The Astrophysical Journal, 668, 417, \dodoi{10.1086/521092}

\bibitem[{{Hasinger}(1987)}]{Hasinger_1987}
{Hasinger}, G. 1987, \aap, 186, 153

\bibitem[{{Hasinger} \& {van der Klis}(1989{\natexlab{a}})}]{Hasinger_Vanderklis_1989}
{Hasinger}, G., \& {van der Klis}, M. 1989{\natexlab{a}}, \aap, 225, 79

\bibitem[{{Hasinger} \& {van der Klis}(1989{\natexlab{b}})}]{1989A&A...225...79H}
---. 1989{\natexlab{b}}, \aap, 225, 79

\bibitem[{Homan {et~al.}(2018)Homan, Steiner, Lin, Fridriksson, Remillard, Miller, \& Ludlam}]{Homan_2018}
Homan, J., Steiner, J.~F., Lin, D., {et~al.} 2018, The Astrophysical Journal, 853, 157, \dodoi{10.3847/1538-4357/aaa439}

\bibitem[{Homan {et~al.}(2002)Homan, van~der Klis, Jonker, Wijnands, Kuulkers, Méndez, \& Lewin}]{Homan_2002}
Homan, J., van~der Klis, M., Jonker, P.~G., {et~al.} 2002, The Astrophysical Journal, 568, 878, \dodoi{10.1086/339057}

\bibitem[{Homan {et~al.}(2007)Homan, van~der Klis, Wijnands, Belloni, Fender, Klein-Wolt, Casella, Méndez, Gallo, Lewin, \& Gehrels}]{Homan_2007}
Homan, J., van~der Klis, M., Wijnands, R., {et~al.} 2007, The Astrophysical Journal, 656, 420, \dodoi{10.1086/510447}

\bibitem[{{in 't Zand} {et~al.}(1999){in 't Zand}, {Verbunt}, {Strohmayer}, {Bazzano}, {Cocchi}, {Heise}, {van Kerkwijk}, {Muller}, {Natalucci}, {Smith}, \& {Ubertini}}]{Rw}
{in 't Zand}, J.~J.~M., {Verbunt}, F., {Strohmayer}, T.~E., {et~al.} 1999, \aap, 345, 100, \dodoi{10.48550/arXiv.astro-ph/9902319}

\bibitem[{Ingram {et~al.}(2009)Ingram, Done, \& Fragile}]{10.1111/j.1745-3933.2009.00693.x}
Ingram, A., Done, C., \& Fragile, P.~C. 2009, Monthly Notices of the Royal Astronomical Society: Letters, 397, L101, \dodoi{10.1111/j.1745-3933.2009.00693.x}

\bibitem[{{Jackson, N. K.} {et~al.}(2009){Jackson, N. K.}, {Church, M. J.}, \& {Bałucińska-Church, M.}}]{refId0}
{Jackson, N. K.}, {Church, M. J.}, \& {Bałucińska-Church, M.} 2009, A\&A, 494, 1059, \dodoi{10.1051/0004-6361:20079234}

\bibitem[{Jonker {et~al.}(2000)Jonker, van~der Klis, Wijnands, Homan, van Paradijs, Méndez, Ford, Kuulkers, \& Lamb}]{Jonker_2000}
Jonker, P.~G., van~der Klis, M., Wijnands, R., {et~al.} 2000, The Astrophysical Journal, 537, 374, \dodoi{10.1086/309029}

\bibitem[{{Kuulkers} {et~al.}(1994){Kuulkers}, {van der Klis}, {Oosterbroek}, {Asai}, {Dotani}, {van Paradijs}, \& {Lewin}}]{Cyg_like}
{Kuulkers}, E., {van der Klis}, M., {Oosterbroek}, T., {et~al.} 1994, \aap, 289, 795

\bibitem[{Lin {et~al.}(2012)Lin, Remillard, Homan, \& Barret}]{Lin_2012}
Lin, D., Remillard, R.~A., Homan, J., \& Barret, D. 2012, The Astrophysical Journal, 756, 34, \dodoi{10.1088/0004-637X/756/1/34}

\bibitem[{{Mitsuda} {et~al.}(1989){Mitsuda}, {Inoue}, {Nakamura}, \& {Tanaka}}]{1989PASJ...41...97M}
{Mitsuda}, K., {Inoue}, H., {Nakamura}, N., \& {Tanaka}, Y. 1989, \pasj, 41, 97

\bibitem[{{Mitsuda} {et~al.}(1984){Mitsuda}, {Inoue}, {Koyama}, {Makishima}, {Matsuoka}, {Ogawara}, {Shibazaki}, {Suzuki}, {Tanaka}, \& {Hirano}}]{1984PASJ...36..741M}
{Mitsuda}, K., {Inoue}, H., {Koyama}, K., {et~al.} 1984, \pasj, 36, 741

\bibitem[{Muno {et~al.}(2003)Muno, Özel, \& Chakrabarty}]{Muno_2003}
Muno, M.~P., Özel, F., \& Chakrabarty, D. 2003, The Astrophysical Journal, 595, 1066, \dodoi{10.1086/377447}

\bibitem[{P \& Sriram(2022)}]{sriram_2022}
P, C., \& Sriram, K. 2022, Monthly Notices of the Royal Astronomical Society, 516, 2500, \dodoi{10.1093/mnras/stac2319}

\bibitem[{Ross {et~al.}(1999)Ross, Fabian, \& Young}]{reflection_ionized}
Ross, R.~R., Fabian, A.~C., \& Young, A.~J. 1999, Monthly Notices of the Royal Astronomical Society, 306, 461, \dodoi{10.1046/j.1365-8711.1999.02528.x}

\bibitem[{{Shakura} \& {Sunyaev}(1973)}]{1973A&A....24..337S}
{Shakura}, N.~I., \& {Sunyaev}, R.~A. 1973, \aap, 24, 337

\bibitem[{{Singh} {et~al.}(2017){Singh}, {Stewart}, {Westergaard}, {Bhattacharayya}, {Chandra}, {Chitnis}, {Dewangan}, {Kothare}, {Mirza}, {Mukerjee}, {Navalkar}, {Shah}, {Abbey}, {Beardmore}, {Kotak}, {Kamble}, {Vishwakarama}, {Pathare}, {Risbud}, {Koyande}, {Stevenson}, {Bicknell}, {Crawford}, {Hansford}, {Peters}, {Sykes}, {Agarwal}, {Sebastian}, {Rajarajan}, {Nagesh}, {Narendra}, {Ramesh}, {Rai}, {Navalgund}, {Sarma}, {Pandiyan}, {Subbarao}, {Gupta}, {Thakkar}, {Singh}, \& {Bajpai}}]{SXT}
{Singh}, K.~P., {Stewart}, G.~C., {Westergaard}, N.~J., {et~al.} 2017, Journal of Astrophysics and Astronomy, 38, 29, \dodoi{10.1007/s12036-017-9448-7}

\bibitem[{{Tauris} \& {van den Heuvel}(2006)}]{2006csxs.book..623T}
{Tauris}, T.~M., \& {van den Heuvel}, E.~P.~J. 2006, in Compact stellar X-ray sources, Vol.~39, 623--665, \dodoi{10.48550/arXiv.astro-ph/0303456}

\bibitem[{{Titarchuk} {et~al.}(2001){Titarchuk}, {Bradshaw}, {Geldzahler}, \& {Fomalont}}]{Titarchuk_2001}
{Titarchuk}, L.~G., {Bradshaw}, C.~F., {Geldzahler}, B.~J., \& {Fomalont}, E.~B. 2001, \apjl, 555, L45, \dodoi{10.1086/323160}

\bibitem[{{van der Klis}(2000{\natexlab{a}})}]{2000ARA&A..38..717V}
{van der Klis}, M. 2000{\natexlab{a}}, \araa, 38, 717, \dodoi{10.1146/annurev.astro.38.1.717}

\bibitem[{{van der Klis}(2000{\natexlab{b}})}]{vanderklis_2000}
---. 2000{\natexlab{b}}, \araa, 38, 717, \dodoi{10.1146/annurev.astro.38.1.717}

\bibitem[{van~der Klis(2004)}]{vanderklis2004review}
van~der Klis, M. 2004, A review of rapid X-ray variability in X-ray binaries.
\newblock \doarXiv{astro-ph/0410551}

\bibitem[{{van der Klis} {et~al.}(1985){van der Klis}, {Jansen}, {van Paradijs}, {Lewin}, {van den Heuvel}, {Trumper}, \& {Szatjno}}]{1985Natur.316..225V}
{van der Klis}, M., {Jansen}, F., {van Paradijs}, J., {et~al.} 1985, \nat, 316, 225, \dodoi{10.1038/316225a0}

\bibitem[{van Paradijs {et~al.}(1988)van Paradijs, Hasinger, Lewin, van~der Klis, Sztajno, Schulz, \& Jansen}]{10.1093/mnras/231.2.379}
van Paradijs, J., Hasinger, G., Lewin, W. H.~G., {et~al.} 1988, Monthly Notices of the Royal Astronomical Society, 231, 379, \dodoi{10.1093/mnras/231.2.379}

\bibitem[{White {et~al.}(1988)White, Stella, \& Parmar}]{White_1988}
White, N.~E., Stella, L., \& Parmar, A.~N. 1988, THE X-RAY SPECTRAL PROPERTIES OF ACCRETION DISKS IN X-RAY BINARIES

\bibitem[{Wijnands {et~al.}(1998)Wijnands, Méndez, van~der Klis, Psaltis, Kuulkers, \& Lamb}]{Wijnands_1998}
Wijnands, R., Méndez, M., van~der Klis, M., {et~al.} 1998, The Astrophysical Journal, 504, L35, \dodoi{10.1086/311564}

\bibitem[{{Wilms} {et~al.}(2000){Wilms}, {Allen}, \& {McCray}}]{2000ApJ...542..914W}
{Wilms}, J., {Allen}, A., \& {McCray}, R. 2000, \apj, 542, 914, \dodoi{10.1086/317016}

\bibitem[{{Zdziarski} {et~al.}(1996){Zdziarski}, {Gierlinski}, {Gondek}, \& {Magdziarz}}]{1996A&AS..120C.553Z}
{Zdziarski}, A.~A., {Gierlinski}, M., {Gondek}, D., \& {Magdziarz}, P. 1996, \aaps, 120, 553

\bibitem[{{Zhang} {et~al.}(1995){Zhang}, {Jahoda}, {Swank}, {Morgan}, \& {Giles}}]{Zang1995}
{Zhang}, W., {Jahoda}, K., {Swank}, J.~H., {Morgan}, E.~H., \& {Giles}, A.~B. 1995, \apj, 449, 930, \dodoi{10.1086/176111}

\bibitem[{{{\.Z}ycki} {et~al.}(1999){{\.Z}ycki}, {Done}, \& {Smith}}]{1999MNRAS.309..561Z}
{{\.Z}ycki}, P.~T., {Done}, C., \& {Smith}, D.~A. 1999, \mnras, 309, 561, \dodoi{10.1046/j.1365-8711.1999.02885.x}

\end{thebibliography}
\bibliographystyle{aasjournal}

\end{document}